\documentclass[10pt,3p,sort&compress]{elsarticle}

\usepackage{amsmath,amssymb}
\usepackage{epsfig}
\usepackage{graphicx}
\usepackage{bm}
\usepackage{amsmath}

\newcommand{\beq}{\begin{equation}}
\newcommand{\eeq}{\end{equation}}
\newcommand{\beqa}{\begin{eqnarray}}
\newcommand{\eeqa}{\end{eqnarray}}
\newcommand{\beqar}{\begin{eqnarray*}}
\newcommand{\eeqar}{\end{eqnarray*}}

\begin{document}

\begin{frontmatter}

\title{ Entropy of an extremal electrically charged thin
shell and the extremal black hole}

\author[jpsl]{Jos\'e P. S. Lemos}
\ead{joselemos@ist.utl.pt}
\address[jpsl]{Centro Multidisciplinar de Astrof\'{\i}sica -- CENTRA,
Departamento de F\'{\i}sica,  Instituto Superior T\'ecnico - IST,
Universidade de Lisboa - UL,
Avenida Rovisco Pais 1, 1049-001 Lisboa, Portugal,\vspace*{.2cm} }
\author[gmq]{Gon\c{c}alo M. Quinta}
\ead{goncalo.quinta@ist.utl.pt}
\address[gmq]{Centro Multidisciplinar de Astrof\'{\i}sica -- CENTRA,
Departamento de F\'{\i}sica,  Instituto Superior T\'ecnico - IST,
Universidade de Lisboa - UL,
Avenida Rovisco Pais 1, 1049-001 Lisboa, Portugal, \vspace*{.2cm} }
\author[obz]{Oleg B. Zaslavskii}
\ead{ozaslav@kharkov.ua}
\address[obz]{Astronomical Institute of Kharkov V.~N.~Karazin National
University, 35
Sumskaya St., Kharkov, 61022, Ukraine, and\\
Institute of
Mathematics and Mechanics, Kazan Federal University, 18 Kremlyovskaya 
Street,
Kazan 420008, Russia.}

\begin{abstract}
There is a debate as to what is the value of the the entropy
$S$ of extremal black holes. There are approaches that yield zero
entropy $S=0$, while there are others that yield the
Bekenstein-Hawking entropy $S=A_+/4$, in Planck units.  There are
still other approaches that give that $S$ is proportional to $r_+$ or
even that $S$ is a generic well-behaved function of $r_+$.  Here $r_+$
is the black hole horizon radius and $A_+=4\pi r_+^2$ is its horizon
area.  Using a spherically symmetric
thin matter shell with extremal electric charge, we
find the entropy expression for the extremal thin shell spacetime.
When the shell's radius approaches its own gravitational radius, and
thus turns into an extremal black hole, we encounter that the entropy
is $S=S(r_+)$, i.e., the entropy of an extremal black hole is a
function of $r_+$ alone.  We speculate that the range of values for an
extremal black hole is $0\leq S(r_+) \leq A_+/4$.
\end{abstract}



\end{frontmatter}

\section{Introduction}

The entropy $S$ and thermodynamics of black holes have been worked out
first by Bekenstein \cite{beken} and Hawking and collaborators
\cite{hawk1,hawk2}.  The Bekenstein-Hawking entropy is given by
$S=A_{+}/4$, where $A_{+}=4\pi r_{+}^{2}$,  $A_{+}$ and $r_{+}$
are the horizon  area and the horizon radius, respectively, and we are
putting all the natural constants equal to one, i.e., we use Planck
units. York and collaborators
\cite{york1,yorketal,yorkmart} (see also
\cite{zaslcanon,pecalemos}) have further worked
out the black hole thermodynamic properties by using canonical and
grand canonical thermodynamic ensembles. There are several other
methods that can be used to study black hole thermodynamics, one that
suits us here uses matter shells
\cite{Mart,lemosquintazaslavskii3,llmr}. In this method, one studies the
generic thermodynamics of the shells at any shell radius, and as one
sends the shell to its own gravitational radius one recovers the $
S=A_{+}/4$ Bekenstein-Hawking entropy. This is the quasiblack hole
method, the evident power of it was displayed in
\cite{lemoszaslavskii}.

A particular class of black holes is the extremal black hole class.
Electrically charged black holes in general
relativity, the ones we are interested here,
have $ m\geq Q$, and the extremal black holes are characterized by
having their mass $m$ equal to their electric charge $Q$, $m=Q$. The
extremal black holes seem to have distinct properties. For instance,
according to the Hawking temperature formula, extremal black holes have
zero temperature. In addition the entropy of an extremal black hole is
a subject of a wide debate as there are different reasonings that can
be applied which lead to different values for the entropy. Hawking and
collaborators \cite{Hawk2} and Teitelboim \cite {Teit} have given
topological arguments which point to the conclusion that extremal
black holes have zero entropy. Further evidence from other arguments
for $S=0$ for extremal black holes was provided in 
\cite{gibbskallosh,ghoshmitra1997,hod}, see also
\cite{cjr,ederyconstantineau}.  One could also argue, naively, that
since the Hawking temperature is zero, then according to one of the
formulations of the third law of thermodynamics as many textbooks
present it should have zero entropy.

However, there remain doubts why the Bekenstein-Hawking formula does
not hold. After all, working out the entropy of non-extremal black
holes and taking the extremal limit $m=Q$ yields $S=A_+/4$, see, e.g.,
\cite {hawk1,hawk2,yorketal,lemosquintazaslavskii3}. In this case, the
thermodynamic argument would not hold, the extremal black hole could
be a system of minimum energy and degenerate ground state and such
systems can have entropy even at zero temperature. Moreover, in string
theory, there are arguments, other than geometrical, that make use of
a direct counting of string and D-brane states in composite systems,
which manipulate carefully the turning on of gravity and electricity
adiabatically by equal amounts to maintain extremality without
changing the counting of states and thus the entropy, that show that
the entropy of an extremal black hole is $S=A_+/4$, as first delivered
by Strominger and Vafa \cite{stromingervafa}, see \cite {sen} for a
review. Other methods also indicate the $S=A_+/4$ value. These are
methods that use quantum field corrections to the black hole entropy
\cite{mannsoloduk}, Hamiltonian methods \cite{kieferlouko,diaslemos},
and thermodynamics ensemble methods
\cite{zaslavskii1,zaslavski1b,zaslavskii5}.  There are works that show
that depending on the black hole type, i.e., black holes with scalar
fields, one has $S=0$ or $S=A_+/4$ \cite {wangsu,wangabdallasu}.

Given that several different calculations give $S=0$ or $S=A_+/4$ one
might open the box and speculate that other values for the entropy are
possible, e.g., any value between 0 and $A_+/4$, or possibly other
functions. Indeed, in a semiclassical calculation of the entropy on an
extremal black hole background Ghosh and Mitra \cite{ghoshmitra1995}
pointed to an entropy value proportional to $r_+$ rather than $A_+$
(i.e., $r_+^2$). This was followed by some discussion for the exact
possible values for the entropy of an extremal black hole
\cite{ghoshmitra1996,zaslavskii2,mitra1998}.

In addition, in a setup using an extremal charged thin shell
contracting reversibly and arranged to maintain extremality it was
shown in \cite{pret} that any value of the entropy of the shell in
passing its own gravitational radius could be achieved. In another
setting, using also thin shells, it was shown that in the quasiblack
hole limit, i.e., when the boundary of the matter is at its own
gravitational radius, the entropy of the extremal black hole is a
generic function of $r_+$ \cite{lemoszaslaextremal}.

One feature important to note is that a calculation of the
stress-energy tensor of quantum fields at the neighborhood of the
event horizon excludes the possibility that an extremal black hole can
be in thermal equilibrium with radiation at any temperature. An
extremal black hole has zero temperature and if the temperature of the
surrounding fields is nonzero then the stress-energy diverges strongly
\cite{andersonhiscock}.

This paper is committed to the study of the entropy of extremally
charged spherically symmetric thin shells of any radii, and in
particular, to the understanding of the entropy of the system when the
radius of the shell is its own gravitational radius, 
i.e., the extremal shell spacetime turns into an 
extremal black hole spacetime. This yields an
expression for the black hole entropy.  
We follow the formalism of
Martinez \cite{Mart} developed for 
electric non-extremal shells in
\cite{lemosquintazaslavskii3}
and for rotating BTZ spacetimes in \cite{llmr}. 
The thermodynamic 
analysis of Callen \cite{callen} is used, as it was used
in \cite
{Mart,lemosquintazaslavskii3,llmr}. The importance of Callen's analysis for
black hole thermodynamics was first understood by York \cite{york1},
see also \cite {yorketal,pecalemos}.
Here, we restrict ourselves to spherically symmetric systems but, as the
results have a rather general character, we believe that, with some
minor changes, they are valid for distorted and rotating systems as
well (see \cite{llmr} for a rotating (2+1)-dimensional
spacetime).

The work is organized as follows: In Sec.~\ref{characteristics} we
give the mechanical properties of an extremal electrically charged thin
shell. Such type of matter is also called electrically counterpoised
dust. In Sec.~\ref {entropyreview} we will analyze the first law of
thermodynamics applied to such a thin shell of any radius and find the
entropy of the spacetime. In Sec.~\ref{bh} we take the shell to its
own gravitational radius and find that the entropy of an extremal
black hole is a generic function of $r_+$. 
We also speculate on the possible values 
for the entropy of an extremal black hole.
In Sec.~\ref{bh2} we display another
interesting shell that can be taken to its own gravitational 
radius and find the corresponding entropy.
In Sec.~\ref{conc} we draw
our conclusions.

\section{The extremal charged thin shell spacetime}

\label{characteristics}

The Einstein-Maxwell equations in four spacetime 
dimensions are given by 
\begin{equation}
G_{\alpha\beta}=8\pi \, T_{\alpha\beta}\,,  \label{ein}
\end{equation}
\begin{equation}
\nabla_{\beta}F^{\alpha\beta}=4\pi J^{\alpha}\,.  \label{max}
\end{equation}
$G_{\alpha\beta}$ is the Einstein tensor, built from the spacetime
metric $ g_{\alpha\beta}$ and its first and second derivatives, $8\pi$
is the coupling, and we are using units in which the velocity of light
is one and the gravitational constant $G$ is also put to one
$G=1$. $T_{\alpha\beta}$ is the energy-momentum
tensor. $F_{\alpha\beta}$ is the Faraday-Maxwell tensor, $J_\alpha$ is
the electromagnetic four-current and $\nabla_\beta$ denotes covariant
derivative. The other Maxwell equation $
\nabla_{[\gamma}F_{\alpha\beta]}=0$, where $[...]$ means
antisymmetrization, is automatically satisfied for a properly defined
$F_{\alpha\beta}$. Greek indices will be used for spacetime indices
and run as $\alpha,\beta=0,1,2,3$, with 0 being the time index.

The concept of thin shell is associated with the presence of matter in
the surface that separate two partitions of spacetime, each with its
own metric.  We will be considering the case of a four-dimensional
spherically symmetric spacetime and a spherical thin shell at some
radius $R$ separating an inner region $\mathcal{V}_{i}$ with flat
metric and an outer region $\mathcal{V}_{o}$ with an extremal
Reissner-Nordstr\"{o}m line element. Thus, for the inner region the
metric is
\begin{align}
ds_{i}^{2}& =g_{\alpha \beta }^{i}dx^{\alpha }dx^{\beta }=  \notag
\label{LEI} \\
& -dt_{i}^{2}+dr^{2}+r^{2}\,d\Omega ^{2}\,,\quad r\leq R\,,
\end{align} 
where $x^{\alpha }=(t_{i},r,\theta ,\phi )$ are the inner coordinates,
with $ t_{i}$ being the inner time, and $(r,\theta ,\phi )$ polar
coordinates, and $ d\Omega ^{2}=d\theta ^{2}+\sin ^{2}\theta \,d\phi
^{2}$. For the outer region the metric is
\begin{align}
ds_{o}^{2}& =g_{\alpha \beta }^{o}dx^{\alpha }dx^{\beta }=  \notag
\label{LEO} \\
& \hspace{-3mm}-\left( 1-\frac{m}{r}\right) ^{2}dt_{o}^{2}+\frac{dr^{2}}{ 
\left( 1-\dfrac{m}{r}\right) ^{2}}  \notag \\
& \hspace{7mm}+r^{2}d\Omega ^{2}\,,\quad r\geq R\,,
\end{align} 
where $x^{\alpha }=(t_{o},r,\theta ,\phi )$ are the outer coordinates,
with $ t_{o}$ being the outer time, and $(r,\theta ,\phi )$ polar
coordinates. In addition, $m$ is the ADM mass, and 
$Q$ is the
electric charge. In the extremal case they are related by
\begin{equation}
m=Q\,.  \label{extremalrel}
\end{equation} 
On the hypersurface itself, $r=R$, the
metric is that of a 2-sphere with an additional time dimension, 
such that the line element is 
\begin{equation}
ds_{\Sigma }^{2}=h_{ab}dy^{a}dy^{b}=-d\tau ^{2}+R^{2}(\tau )d\Omega
^{2}\,,\quad r=R\,,  \label{intrinsmetr}
\end{equation} 
where we have chosen $y^{a}=(\tau ,\theta ,\phi )$ as the time and
spatial coordinates on the shell. Latin indices apply for the
components on the hypersurface. The time coordinate $\tau $ is the
proper time for an observer located at the shell. The shell radius is
given by the parametric equation $R=R(\tau )$ for an observer on the
shell.  We consider a static shell so that $R(\tau)={\rm constant}$.
On each side of the hypersurface, the parametric equations for the
time and radial coordinates are denoted by $ t_{i}=T_{i}(\tau )$,
$r_{i}=R_{i}(\tau )$, and $t_{o}=T_{o}(\tau )$, $ r_{o}=R_{o}(\tau )$.

Imposing that the fluid in the shell is a perfect 
fluid with stress-energy
tensor $S^{a}{}_{b}$ given by 
\begin{equation}
S^{a}{}_{b} = (\sigma +p) u^a u_b + p h^{a}{}_{b}\,,  \label{perffluid}
\end{equation}
where $u^a$ is the 3-velocity of a shell element, one finds 
through the junction conditions (see, e.g., \cite 
{lemosquintazaslavskii3}) 
\begin{align}
\sigma = & \frac{m}{4\pi R^2} \,,  \label{SS1} \\
p = & 0 \,.  \label{SS2}
\end{align}
Matter for which $p=0$ and
totally supported by electric forces 
against gravitational collapse
is called extremal matter or, sometimes, electrically
counterpoised dust.
The rest mass of the shell $M$ is defined as 
\begin{equation}
\sigma = \frac{M}{4\pi R^2}\,,  \label{SS1restmass}
\end{equation}
and so in the extremal case 
\begin{equation}
M=m\,.  \label{admmassrestmass}
\end{equation}

The gravitational radius $r_+$ of the shell
is given by the zero of the  $g_{00}^o$ in Eq.~(\ref{LEO}).
It is actually a double zero, one gives the 
gravitational radius $r_+$, the other the 
Cauchy horizon $r_-$ of the shell. The double zero 
means that for the
extremal 
spacetime  
the two radii 
coincide, 
\begin{equation}
r_+= r_-\,,  \label{zeror+=r-}
\end{equation}
and we call it $r_+$
from now on. The zero of the $g_{00}^o$ in Eq.~(\ref{LEO})
then gives 
\begin{equation}
r_+= m\,,  \label{zeror+}
\end{equation}
and so 
\begin{equation}
r_+= r_-=m=Q=M\,.  \label{zeror+general}
\end{equation}
The gravitational radius $r_+$ is also the horizon radius when the
shell radius $R$ is inside $r_+$, i.e., when the spacetime contains a
black hole.  Clearly, gravitational and horizon radii have the same
expression.  Nevertheless, they are distinct. The gravitational radius
is a property of the matter and the spacetime generated by it,
independently of whether there is a black hole or not, whereas there
is a horizon radius in the case there is a black hole.

One can define the gravitational area $A_+$
as the area spanned by the gravitational radius, namely,
\begin{equation}  \label{arear+}
A_+=4\pi\,r_+^2\,.
\end{equation}
This is also the event horizon area when there is a black hole. 
It is useful
to define the shell's redshift function $k$ as 
\begin{equation}  \label{red}
k=1-\frac{r_+}{R}\,.
\end{equation}
The area $A$ of the shell, an important quantity, is given by 
\begin{equation}
A=4 \pi R^2.  \label{area1}
\end{equation}
Also for a shell with electric charge density $\sigma_e$ one finds 
(see, e.g., \cite 
{lemosquintazaslavskii3}) 
\begin{equation}  \label{PhiJC}
\frac{Q}{R^2}=4\pi \sigma_e \,.
\end{equation}
This equation relates the total charge $Q$, the charge density
$\sigma_e$, and the shell's radius $R$.

An obvious inequality, for the shell is that it should be always
outside its own gravitational radius, so
\begin{equation}
R \geq r_+\,.  \label{notrapped}
\end{equation}
Then the physical allowed values for $k$ in Eq.~(\ref{red}) are in the
interval $[0,1]$. Since the pressure of the matter in the shell is
zero and the energy density is considered positive, see
Eq.~(\ref{SS2}), the energy conditions, weak, strong, and dominant,
are always obeyed for $R \geq r_+$.

It is worth noting that in the limit $R\to r_+$ there are subtleties
connected with the behavior of the boundary's geometry. 
Indeed, there is a discontinuity because of the timelike
character of the boundary from the inside and the lightlike character
of the boundary from the outside (see \cite{lemoszaslavskii} for
details). However, here they are essentially irrelevant since in what
follows we consider the external region only.

\section{Entropy of an extremal charged thin shell}

\label{entropyreview}

\subsection{Entropy and the first law of thermodynamics for an extremal
charged thin shell}

\label{E}

In the study of the thermodynamics and entropy of a thin shell we use
units in which the Boltzmann constant is one. We assume that the shell
in static equilibrium at radius $R$ has a well defined local
temperature $T$ and an entropy $S$. The entropy $S$ is a function of
the shell's rest mass $M$, area $A$, and charge $Q$, i.e.,
\begin{equation}  \label{entropy0}
S=S(M,A,Q)\,.
\end{equation}
The first law of thermodynamics can be then written as 
\begin{equation}  \label{TQ0}
T dS = dM + pdA - \Phi dQ\,,
\end{equation}
or, defining the inverse temperature $\beta$, 
\begin{equation}  \label{beta}
\beta \equiv \frac1T
\end{equation}
one has 
\begin{equation}  \label{TQ}
dS = \beta\left( dM + pdA - \Phi dQ\right)\,,
\end{equation}
where $dS$, $dM$, $dA$, and $dQ$ are the differentials of the entropy,
rest mass, area, and electric charge of the shell, respectively,
whereas $T$, $p$ and $\Phi$ are its temperature, pressure, and
thermodynamic electric potential, respectively. To obtain the entropy
$S$ of the shell, one thus needs in general
to specify three equations of state,
namely, $p=p(M,A,Q)$, $ \beta=\beta(M,A,Q)$, and
$\Phi=\Phi(M,A,Q)$. 

The extremal case is a special case, the extremality condition will
constraint the possible configurations. From Eq.~(\ref{zeror+general}) we
have for an extremal shell 
\begin{equation}  \label{equalQM}
dQ=dM\,.
\end{equation}
Thus, the number of independent variables reduces to two, namely, $M$
and $A$, and so, $p=p(M,A)$, $\beta=\beta(M,A)$, and
$\Phi=\Phi(M,A)$. It is more convenient to work out with the shell's
radius $R$ than its area $A$, which can be done from
Eq.~(\ref{area1}), so the equations of state are of the form
\begin{equation}  \label{eqsstate}
p=p(M,R)\,,\; \beta=\beta(M,R)\,,\; \Phi=\Phi(M,R)\,.
\end{equation}
Now, from Eq.~(\ref{SS2}), one has that the equation 
of state for the pressure is
\begin{equation}  \label{presseqsstate}
p(M,R)=0\,.
\end{equation}
Thus, the the first law (\ref{TQ}) is now 
\begin{equation}  \label{secondlawextremal0}
dS = \beta\left(1-\Phi\right)dM\,,
\end{equation}
and since from 
Eq.~(\ref{zeror+general}) $M=r_+$ and $dM=dr_+$, one can write 
the first law as 
\begin{equation}  \label{secondlawextremal}
dS = \beta\left(1-\Phi\right)dr_+\,,
\end{equation}
where now 
\begin{equation}  \label{eqsstate2}
\beta=\beta(r_+,R)\,,\; \Phi=\Phi(r_+,R)\,.
\end{equation}
The integrability condition for Eq.~(\ref{secondlawextremal})
reduces to a simple equation, namely, 
\begin{equation}  \label{integextr}
\beta\left(1-\Phi\right)=s(r_+)\,,
\end{equation}
where $s$ is a function of $r_+$ alone and is arbitrary as long as it
gives a positive meaningful entropy. 
Since $\beta\geq0$ and $s\geq0$ we have the 
following constraint on $\Phi$,
\begin{equation}  \label{constraintPhi}
\Phi\leq1\,.
\end{equation}
The result given in Eq.~(\ref{integextr}), that the most
general function of the product of two functions (namely, 
$\beta$ and
$1-\Phi$) of $r_+$ and $R$ is a function of $r_+$ alone, is new and
interesting. Then Eq.~(\ref{secondlawextremal})
together with Eq.~(\ref{integextr}) yields
\begin{equation}  \label{dSQ2}
dS = s(r_+) dr_+ \,.
\end{equation}
The function $s(r_+)$ is thus a kind of entropy density.
Integrating Eq.~(\ref{dSQ2}), we conclude that the 
entropy of the extremal
shell is given by 
\begin{equation}  \label{S7}
S= S(r_+)\,,\quad R\geq r_+\,,
\end{equation}
where we have assumed that the constant of integration is zero. Thus
the entropy of an extremal charged thin shell is a function of $r_+$
alone.  Depending on the choice of $s(r_+)$ we can obtain a wide range
of values for the entropy $S(r_+)$ of the shell. Since $\beta(r_+,R)$
and $\Phi(r_+,R)$ are arbitrary as long as they obey the constraint
(\ref{integextr}), this shows that the extremal case is indeed quite
special. Such a result does not appear in the non-extremal case at
all, see \cite{lemosquintazaslavskii3}.

\subsection{Choices for the matter equations of state of an extremal
charged thin shell}
\label{E3}

The functions $\beta(r_+,R)$ and $\Phi(r_+,R)$ being arbitrary can be
chosen at our will, bearing in mind Eq.~(\ref{integextr}).

Now, in gravitational thermodynamics, there is the well-known Tolman
formula for static spacetimes. The formula relates the temperature at
$r$ of something, i.e., $T(r)$ with the temperature of that something
at infinity $T_\infty$.  In terms of the inverse temperatures
$\beta(r)=1/T(r)$ and $b=1/T_\infty$ the Tolman formula is given by $
\beta(r)=b\sqrt{-g_{00}(r)}\label{tol}$, where $g_{00}(r)$ is the zero
zero component of the static metric. The quantity $b$ here is a
number, a constant.  Thus, the inverse temperature $\beta(r)$ is
redshifted from $b$, i.e., conversely, the temperature at $r$, $T(r)$,
is blueshifted from the temperature at infinity, $T_\infty$.

Remarkably, for the thin shell, 
the Tolman formula can be derived directly 
through 
the integrability
conditions for 
non-extremal thin shells
(see  \cite{Mart} for neutral shells, and 
\cite{lemosquintazaslavskii3}
for electrically charged  non-extremal shells). 
This is an important particular 
case of the Tolman formula  for which
$r=R$. For generic non-extremal shells 
the integrability conditions give 
$\beta(r_{+},r_-,R)=b(r_{+},r_-)\,k(r_{+},r_-,R)$,
where, 
$k(r_{+},r_-,R)=\sqrt{-g_{00}(R)}$, 
i.e., $k(r_{+},r_-,R)=\sqrt{\left(1-\frac{r_{+}}{R}\right)
\left(1-\frac{r-}{R}\right)}\,$,
see \cite{lemosquintazaslavskii3} for details.
This is also the expression for $\beta$ obtained 
for a nonextremal Reissner-Nordstr\"om black hole, 
see \cite{yorketal}.
Taking the limit to extremal shells, i.e., $r_+=r_-$,
we find
$
\beta(r_{+},R)=b(r_{+})\,k(r_{+},R)\,,
$
where $k(r_{+},R)$ is given in 
Eq.~(\ref{red}).

Now, in the extremal case, the only integrability condition is 
Eq.~(\ref{integextr}), and it has nothing to do with the Tolman formula. 
However, among all other possible choices, Tolman 
formula  $\beta(r_{+},r_-,R)=b(r_{+},r_-)\,k(r_{+},r_-,R)$,
allows for a nontrivial generalization.
It is the Toman formula 
for  extremal shells and 
there is an important difference between the Tolman formula 
for nonextremal
and extremal shells.
For nonextremal shells, one finds from the integrability conditions
that
$b=b(r_{+},r_-)$, i.e., 
$b$ cannot depend on $R$,
see \cite{lemosquintazaslavskii3}. 
For extremal shells, on the other hand, 
nothing prevents us from including in $b$ 
a dependence not only on $r_{+}=r_-$, but
also on $R$, $b=b(r_{+},R)$. As a result the generic Tolman 
formula in the extremal case is then
\begin{equation}
\beta(r_{+},R)=b(r_{+},R)\, k\,,
\label{tolmanextremalpostulated}
\end{equation}
where $k=k(r_{+},R)$ 
as given in Eq.~(\ref{red}).
The function $b(r_+,R\to\infty)$
represents the inverse of the temperature of the shell if it
were located at infinity.
Of course, we can 
invoke the Tolman formula directly from general 
thermodynamic argumentation.

With the choice for $\beta$
given in Eq.~(\ref{tolmanextremalpostulated}), one finds that 
Eq.~(\ref{integextr}) yields
$\Phi(r_+,R) = \frac{
\left(1-\frac{s(r_+)}{b(r_+,R)}\right) - \frac{r_+} {R}}{k}$.
This can be put in the form 
\begin{equation}  \label{Phi0}
\Phi(r_+,R) = \frac{\phi(r_+,R) - \frac{r_+} {R}}{k}\,,
\end{equation}
where we have defined 
$\phi(r_+,R)\equiv 1-\frac{s(r_+)}{b(r_+,R)}$, i.e., 
$\phi$ is such that
\begin{equation}  \label{integextrbphi}
b\left(1-\phi\right)=s(r_+)\,.
\end{equation}
From Eq.~(\ref{Phi0}) one sees that  
$\phi(r_+,R\to\infty)$ represents the electric
potential of the shell if it were located at infinity.

We could proceed and give specific equations
for $b(r_+,R)$ and $\phi(r_+,R)$, and determine the thermodynamic
properties of the shell including its thermodynamic stability.  We
will not do it here, see \cite{Mart,lemosquintazaslavskii3} for some
cases dealing with uncharged shells and non-extremal charged shells,
respectively.
We could also give other equations of state, as long as 
they obey Eq.~(\ref{integextr}).

We stick to the above, 
namely, Eqs.~(\ref{tolmanextremalpostulated})
and (\ref{Phi0}),
and study some particular instances 
that allow us to take the gravitational radius, i.e., 
the black hole, limit.

\section{Entropy of an extremal black hole}

\label{bh}

We want now to study the extremal 
black hole limit, i.e., the case where the
extremal shell is taken to its own gravitational radius, $R=r_+$.

If
we take the extremal shell to its gravitational radius $R=r_{+}$,
i.e., when taking the black hole limit,  we 
have to make several choices. 
First we fix the shell at some radius $R>r_{+}$ and 
choose the functions $\beta$ and $\Phi$, 
or $b$ and $\phi$, appropriately,
only afterwards we send the shell to $R=r_{+}$.
Second, knowing that the Hawking temperature measured at infinity for
an extremal black hole is $T_H=0$, we choose the temperature at
infinity $T_\infty$ as $T_\infty=0$, i.e., $b=\infty$.
Thus from 
Eq.~(\ref{tolmanextremalpostulated})
we find $\beta=\infty$ and so the temperature on 
the shell is zero, $T=0$.
Third we find $\phi$ and $\Phi$. From
Eq.~(\ref{integextrbphi}) we find that
$\phi=1$, such that $1-\phi=s/b$, for some 
well specified $s$. Then also $\Phi=1$.
In other words, 
we can choose $\phi$ as close to 1 as we like
and $T_\infty$ as small as
we want
(or $b$ as large as we want). Afterwards, keeping the product fixed 
(for a given $r_+$), we
can send $\phi$ to 1 and $T_\infty$
to zero (or $b$ to infinity).
We see that the shell at $R>r_+$
has been prepared with $T_\infty=0$ and 
$\phi=1$, such that $b(1-\phi)=s$ and so
$\beta(1-\Phi)=s$. 
The shell is now correctly prepared. 
Having correctly prepared the shell we can
take it to its gravitational radius if we wish.

Now, indeed 
take the shell to its own gravitational radius $R=r_+$,
i.e., take the black hole limit.
Since the entropy differential for the shell
depends only on $r_+$ through the
function $s(r_+)$ that is arbitrary, see
Eq.~(\ref{dSQ2}),
we conclude that the entropy of
the extremal shell in the extremal black hole limit is given by
\begin{equation}  \label{S7777}
S=S(r_+) \,,\quad R= r_+.
\end{equation}
This is the extremal black hole limit 
of an extremal shell. Such a configuration is a quasiblack hole.

Our approach implies that
the entropy of an extremal black hole can assume any well-behaved
function of $r_+$. The precise function of the entropy depends on the
constitution of the matter that collapsed to form the black
hole. Depending on the choice of $s$ that, in turn, depends on 
the choices for $\beta$ and
$\Phi$, we can obtain any function of $r_+$ for the entropy $S$ of the
extremal black hole.  The fact that the entropy in the extremal case
is model-dependent agrees with the previous work
\cite{lemoszaslaextremal} and more early studies \cite{pret}.
Of course, a particular class of entropies for the extremal
black hole would be the Bekenstein-Hawking entropy $S(r_+)=\frac14
A_+=\pi r_+^2$ in units where Planck's constant, along with the
velocity of light and $G$, is also put to one, i.e., in Planck units.
In summary, 
our result is quite different from the non-extremal case, where the
entropy can only have the Bekenstein-Hawking functional dependence
$S(r_{+})=\frac{1}{4}A_{+}$ \cite{lemosquintazaslavskii3}.

Note, that, although the importance of the product $\beta \left( 1-\Phi
\right)$ has been raised in the 
extremal black hole context in \cite{ghoshmitra1995}
(see also \cite{ghoshmitra1996}), the result that the most general function
of the product $\beta \left( 1-\Phi \right) $ is a well-behaved,
but otherwise arbitrary, function of $r_{+}$ is
new. There are additional differences between
\cite{ghoshmitra1995,ghoshmitra1996} and our work. In
\cite{ghoshmitra1995,ghoshmitra1996}, the product $ \beta
(1-\Phi)$ enters the path integral over fluctuating geometries, so it
appears in a quantum context. In doing so, finite nonzero $\beta $
are not forbidden.
However, for such $\beta$ quantum backreaction
destroys the extremal horizon \cite{andersonhiscock}. In our approach,
we consider a shell, not a black hole, and can adjust $\beta $ 
and $\Phi $
at the shell radius in such a way that for any $R$ close to $r_{+}$
backreaction remains finite \cite{lemoszaslaextremal}.

Our preceding calculations and discussion were
rigorous.  Now, we speculate
on ways to constrain the entropy function $S(r_+)$
for the extremal black hole.
For
instance, the 
initial Bekenstein arguments for black holes \cite {beken}, 
non-extremal ones,
proved that an entropy proportional to $A_{+}^{1/2}$ should be
discarded on the basis of the second law of 
thermodynamics. However, since extremal black holes have a 
different character
from non-extremal ones, these arguments do not hold here. Another 
possible
constraint is the following. For the usual, non-extremal, 
black holes the entropy is 
$S(r_{+})=\frac{1}{4}A_{+}$. In this case, when one takes the shell to its
own gravitational radius the pressure at the shell blows up, $p\rightarrow
\infty $ \cite{lemosquintazaslavskii3}, and the spacetime 
is assumed to take the Hawking
temperature. In a sense this means that all possible degrees of freedom are
excited and the black hole takes the Bekenstein-Hawking entropy, 
the maximum possible entropy. Taking the extremal limit
from a non-extremal black hole, one finds 
that in this particular limit the extremal black hole
entropy is the Bekenstein-Hawking entropy.
Thus, this suggests that 
the maximum entropy that an extremal black hole
can take is the Bekenstein-Hawking entropy. 
Therefore, in this sense,
the range of values for the entropy of an
extremal black hole is 
\begin{equation}  \label{Srange}
0\leq S(r_{+})\leq \frac{1}{4}A_{+}\,,
\end{equation}
or $0\leq S(r_{+})\leq \pi r_+^2$.
The case studied by Ghosh and Mitra
\cite{ghoshmitra1995,ghoshmitra1996} has $S\propto r_{+}$ and so is
within our limits.

Table 1 below summarizes the comparison between 
an extremal shell  at its own
gravitational radius with $T_\infty=0$, which we
have called a special shell, 
and an extremal black hole with $T_\infty=0$.

\vskip 0.2cm
{\hskip -0.5cm
\begin{tabular}
[c]{|l|l|l|l|l|l|}\hline
Case & $T$ at $\infty$ & Local $T$ on $r_+$ & 
Backreaction on $r_+$ & $\Phi$  &
Entropy\\\hline
Special shell & 0 & $0$ & Finite & 
$1$ on 
$R$ & 
Well-defined
$S(r_+)$\\\hline
Black hole & 0 & $0$  & Finite & $1$ on $r_+$ & 
In debate,
$S=0$, $\frac{A_+}{4}$, $S(r_+)$\\\hline
\end{tabular}
}
\vskip -0.1cm
\noindent 
Table 1. Comparison between 
a special extremal shell at its own
gravitational radius 
and an extremal black hole both with $T_\infty=0$.

\section{Another interesting shell at the gravitational 
radius limit: A generic shell}

\label{bh2}

There is a more generic shell that 
has 
interesting properties and can be taken also 
to the gravitational radius limit with no 
unbound back reaction.

We now suppose that the shell 
has a small nonzero local temperature (i.e., finite large
$\beta$), rather than zero.
We follow  Eq.~(\ref{tolmanextremalpostulated})
and Eq.~(\ref{Phi0})
and keep in mind
the constraints (\ref{constraintPhi}) and
(\ref{integextrbphi}).

From Eq.~(\ref{tolmanextremalpostulated}) we see that 
the product $bk$ is the important quantity.
At any $R$ prepare the shell so that 
$b={\bar b}/k$, i.e., 
$b={\bar b}/(1-r_+/R)$, for some ${\bar b}$ finite.
Then $\beta={\bar b}$ and is finite, and holds
for any $R>r_+$.
Note that the temperature 
measured at infinity $T_\infty=1/b$ is finite.
Prepare also the shell so that
$(1-\phi)=(1-\bar\phi)k$, so that
$(1-\Phi)=(1-\bar\phi)$
for any $R>r_+$. Note that 
the potential measured at infinity 
$\phi$ is less than one. 

Now, take the gravitational radius limit, $R=r_+$.
In this limit $k$ goes to
zero, but we have prepared the shell  to 
keep $\beta$ bounded even in this black hole limit,
as a small $k$ has been compensated by a large $b$. 
In this limit the temperature 
measured at infinity $T_\infty=1/b$ is zero
and thus
coincides with the Hawking temperature, 
$T_{H}=0$. The temperature $T=1/\beta$
at the shell is nonzero and finite. 
Thus in this case,
quantum backreaction also remains bounded
even for $R=r_{+}$. 
Since $\beta$ is finite and
we have $\beta(1-\Phi)=s(r_+)$ 
we have $\Phi<1$.
The entropy of the shell at $R=r_+$ is 
then also $S=S(r_+)$.
It is
this way of reasoning that was used in \cite{lemoszaslavskii} in
a general discussion of the entropy for the extremal case. In doing so,
any $\beta$ and $\Phi<1$ obeying Eq.~(\ref{integextr}) are suitable,
and an entropy as an arbitrary
(within limits of physical 
reasonability) function of
$r_{+}$, $S=S(r_+)$ was obtained.
This case sharply contrasts with the extremal black hole case where
any $T_\infty=b^{-1}\neq0$ leads to $\beta\rightarrow0$ and infinite
local temperature $\beta^{-1}$ on the horizon with divergent
backreaction that destroys the horizon (for this extremal black hole
case see \cite{Hawk2} where nothing is said about quantum backreaction
and it is argued that the entropy is zero, however results in
\cite{andersonhiscock} show that the backreaction grows unbound if
$T_H$ is not zero).

Table 2 below summarizes the comparison between 
an extremal shell  at its own
gravitational radius with $T_\infty=0$ and 
$\beta$ finite nonzero, which we
have called a generic shell, 
and an extremal black hole with $T_\infty$
not zero. 
\vskip 0.1cm
{\hskip -0.5cm
\begin{tabular}
[c]{|l|l|l|l|l|l|}\hline
Case & $T$ at $\infty$ & Local $T$ on $r_+$ & 
Backreaction on $r_+$ & $\Phi$  & Entropy\\\hline
Generic shell & 0 & Finite, not 0 & Finite & $<1$ 
on $R$ & Well-defined $S(r_+)$\\\hline
Black hole & $T_\infty\neq0$ & $\infty$ & Infinite & 
$1$ & Not known or undefined\\\hline
\end{tabular}
}
\vskip -0.1cm
\noindent 
Table 2. Comparison between 
a generic  extremal shell at its own
gravitational radius 
and an extremal black hole with infinite temperature at
the horizon.
\vskip 0.2cm

\section{Conclusions}

\label{conc}

Upon consideration of spherically symmetric
systems and through the formalism
of thin matter shells and their thermodynamics properties,
we have shown our solution for the ongoing debate concerning the
entropy of an extremal black hole.
Although it would be necessary a full quantum
theory of gravity to fully understand the result obtained, it is
nonetheless interesting to see that the use of the junction conditions
for Einstein equation
leads inevitably to the suggestion that extremal black holes are a
different class of objects than non-extremal black holes, due to the
fact that their entropy depends on the particularities of the matter
distribution which originated the black hole.

\section*{Acknowledgments}

We thank FCT-Portugal for financial support through Project
No.~PEst-OE/FIS/UI0099/2014. GQ 
also ack\-now\-ledges the grant No.~SFRH/BD/92583/2014 from FCT.
OBZ has been partially supported by the Kazan Federal University 
through a state grant for scientific activities.

\end{document}